\documentclass[twocolumn,showpacs,preprintnumbers,amsmath,amssymb]{revtex4}

\usepackage{graphicx}
\usepackage{dcolumn}
\usepackage{bm}

\begin{document}
\title{Heavy-tailed statistics in short-message communication}
\author{Wei Hong$^1$}
\author{Xiao-Pu Han$^1$}
\email{hxp@mail.ustc.edu.cn}
\author{Tao Zhou$^{1, 2}$}
\email{zhutou@ustc.edu(corresponding-author)}
\author{Bing-Hong Wang$^{1, 3}$}
\email{bhwang@ustc.edu.cn}

\affiliation{$^1$Department of Modern Physics, University of Science
and Technology of China, Hefei 230026, P. R. China\\$^2$Department
of Physics, University of Fribourg, CH-1700 Fribourg, Switzerland\\
$^3$Shanghai Institute for Systemic Sciences, Shanghai 200093, P. R.
China}
\date{\today}

\begin{abstract}

Short-message (SM) is one of the most frequently used communication
channels in the modern society. In this Brief Report, based on the
SM communication records provided by some volunteers, we investigate
the statistics of SM communication pattern, including the interevent
time distributions between two consecutive short messages and two
conversations, and the distribution of message number contained by a
complete conversation. In the individual level, the current
empirical data raises a strong evidence that the human activity
pattern, exhibiting a heavy-tailed interevent time distribution, is
driven by a non-Poisson nature.
\end{abstract}

\pacs{89.75.Da, 02.50.-r}
\maketitle

\section{INTRODUCTION}
Information communication builds the basis of social relations. In
the modern science, statistical analysis on the communication
database becomes one of the most important approaches to reveal the
social structure \cite{Watts2007}. For example, the communication
structures based on E-mail \cite{r1} and phone-call \cite{r2}
display scale-free and small-world properties, which are ubiquitous
in various social networks. For the lack of long-term standard
database about human communication activities, prior studies on
social communication systems usually simply assume the temporal
occurrence of contacts between two people is uniform. That is to
say, given two nodes in a social acquaintance network, at any time
the occurring probability of a new contact (e.g. telephone call, SM
communication, on-line instant chat, e-mail communication, etc.) is
the same, which leads to a Poisson distribution of interevent time
between two consecutive contacts. However, recently, the empirical
investigations on e-mail \cite{r18} and surface mail \cite{r19}
communication show a far different scenario: those communication
patterns follow non-Poisson statistics, characterized by bursts of
rapidly occurring events separated by long gaps. That is, the
interevent time distribution has a much fatter tail than the
exponential form, approximated to a power law. The similar
statistical properties have also been found in many other human
behaviors \cite{Zhou2008}, including market transaction
\cite{r20,r21}, on-line game playing \cite{r22}, movie watching
\cite{r23}, web browsing \cite{r24}, and so on. Those empirical
statistics clearly indicate the invalidity of Poisson process in
mimicking the human dynamics in many real-life systems. Motivated by
those empirical evidences, scientists are desired to uncover the
origin of heavy-tails in human dynamics, as well as to reveal the
effect of non-Poisson statistics on some dynamical process in social
systems. In the former aspect, Barab\'asi \emph{et al.} suggested
the highest-priority-first (HPF) protocol a potential origin
\cite{r18,r25}, however, this queuing model may not well explain all
the possible mechanisms leading to a heavy tail \cite{r26}. In the
latter aspect, so far, only few works about epidemic spreading are
reported \cite{r27}.

Based on the rapid progress on wireless techniques, the SM
communication becomes one of the most important social contact tools
in the modern society. In this Brief Report, based on the SM
communication records provided by some volunteers, we investigate
the statistics of SM communication pattern, including the interevent
time distributions between two consecutive short messages and two
conversations, and the distribution of message number contained by a
complete conversation. In the individual level, the current
empirical data raises a strong evidence that the human activity
pattern, exhibiting a heavy-tailed interevent time distribution, is
driven by a non-Poisson nature.

\begin{figure}
  \includegraphics[width=9.0cm]{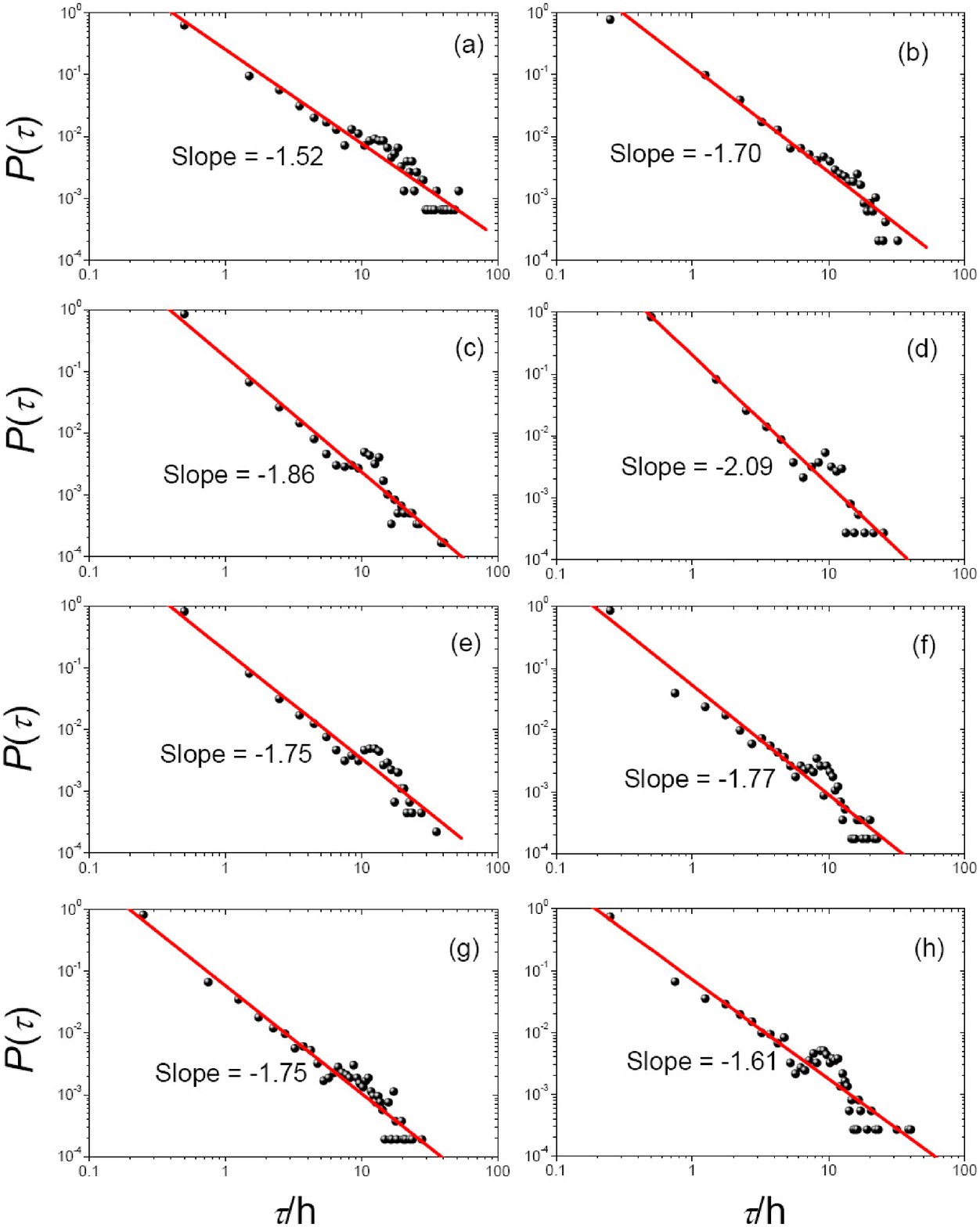}\\
  \caption{(color online)
  The log-log plots of interevent time distributions. The black dots represent the empirical data, and the red lines
  are the linear fittings. The panels (a)-(h) correspond to the records of A-H, respectively.}
\end{figure}

\section{Data}
In this Brief Report, the SM records of eight volunteers (signed
from A to H) are investigated. Those contain one company manager (C)
and seven University students (else). The overall time spans of
those records range from three to six months. Every record contains
the sending time of SMs, and the records A, C, E, F contain the cell
phone numbers of the receivers. Some basic properties of the records
are listed in the Table 1.

\begin{table}[!h]
\tabcolsep 0pt \caption{Properties of the records provided by the
volunteers.} \vspace*{-20pt}
\begin{center}
\def\temptablewidth{0.50\textwidth}
{\rule{\temptablewidth}{1pt}}
\begin{tabular*}{\temptablewidth}{@{\extracolsep{\fill}}ccccccc}
Records & Time span & Total number & Average
number of \\
 & (month) &  of SMs & SMs sent per day \\ \hline
       A  & 6  & 1528  &  8.7\\
       B  & 6  & 4844  &  26.8\\
       C  & 6  & 5987  &  33.6 \\
       D  & 3  & 3780  &  41.1 \\
       E  & 6  & 4523 (2263\footnote{Number of SMs recording the cell phone numbers of receivers.})  &  25.4\\
       F  & 4.5 & 5778  &  41.6\\
       G  & 5  & 5346  &  34.9\\
       H  & 5.5 & 3734  &  22.1\\
       \end{tabular*}
       {\rule{\temptablewidth}{1pt}}
       \end{center}
\end{table}

We focus on the common properties shared by different records, for
those may imply some general statistical characteristics of the
human temporal activities. The time is coarse-grained in an hour
resolution.

\section{Empirical results}
In this section, we present the empirical results of (i) the
interevent time distribution of sending SMs, (ii) the distribution
of interevent time between two consecutive conversations, and (iii)
the length distribution of conversations. The results indicate that
the communication behaviors of different users share some common
properties, especially the observed heavy tails in all those
distributions.

\begin{figure}
  \includegraphics[width=7.0cm]{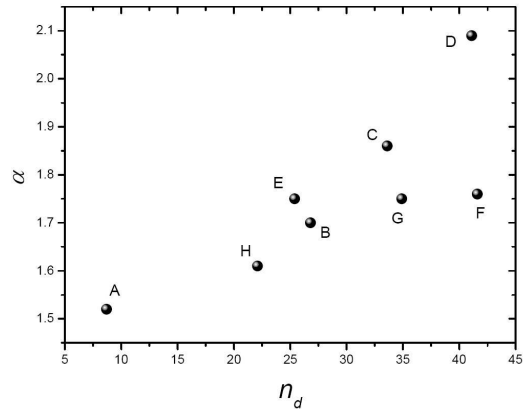}\\
  \caption{Dependence between the exponent $\alpha$ of each distribution and
its daily average number, $n_{d}$.}
\end{figure}

As shown in Fig. 1, the interevent distribution can be well fitted
by a power-law form $P(\tau)\varpropto \tau^{-\alpha}$, where the
exponent $\alpha$ is between 1.5 and 2.1 for different users. In
each curve, an obvious peak turns up at about ten hours, which is
relative to the physiological period of human: the sleeping time are
more or less ten hours. Besides, as shown in Fig. 2, there are
apparently positive correlation between the average numbers of SMs
sent per day and the exponent $\alpha$. This phenomenon, similar to
the observations in the on-line movie watching \cite{r23}, is
against the hypothesis \cite{r25} on the discrete universality
classes of human dynamics.

Generally speaking, in the SM communication, a user often
consecutively sends/receives messages to/from one person. According
to our daily experience, when people are engaged in the SM
communication, they often need several times of exchanging messages
to build a complete conversation. Therefore, we define a
conversation as SMs that are sent to one person consecutively,
without being interrupted by some other persons. Furthermore, the
interevent time between two consecutive conversations is defined as
the time difference between the two beginning times. Only the
samples A, C, E and F are investigated since their records contain
the phone numbers of receivers. As shown in Fig. 3, the interevent
time distributions of conversations can be well fitted by power
laws, and the values of exponents are less than the corresponding
distributions of SM sending. Actually, the conversation rather than
a single SM can better characterize the communication pattern
because a conversation is functional complete. Yet it should be
reminded that, in the real communication process, one complete
conversation may be interrupted by other correspondents, so that a
single conversation may be regarded as several small ones. Since
these conditions cannot be automatically discriminated from the
empirical data, they may cause some bias to a certain extent.
However, the statistical characteristics such as heavy tails shown
in Fig. 3 are believable, because the range of the distribution will
be enlarged and the tail of the distribution would be even fatter if
the bias was eliminated. To characterize the strength of a
conversation, we define a length of a conversation as the number of
SMs it contains. As shown in Fig. 4, every length distribution is
dominated by a power law with exponent larger than 2.

\begin{figure}
  \includegraphics[width=8.5cm]{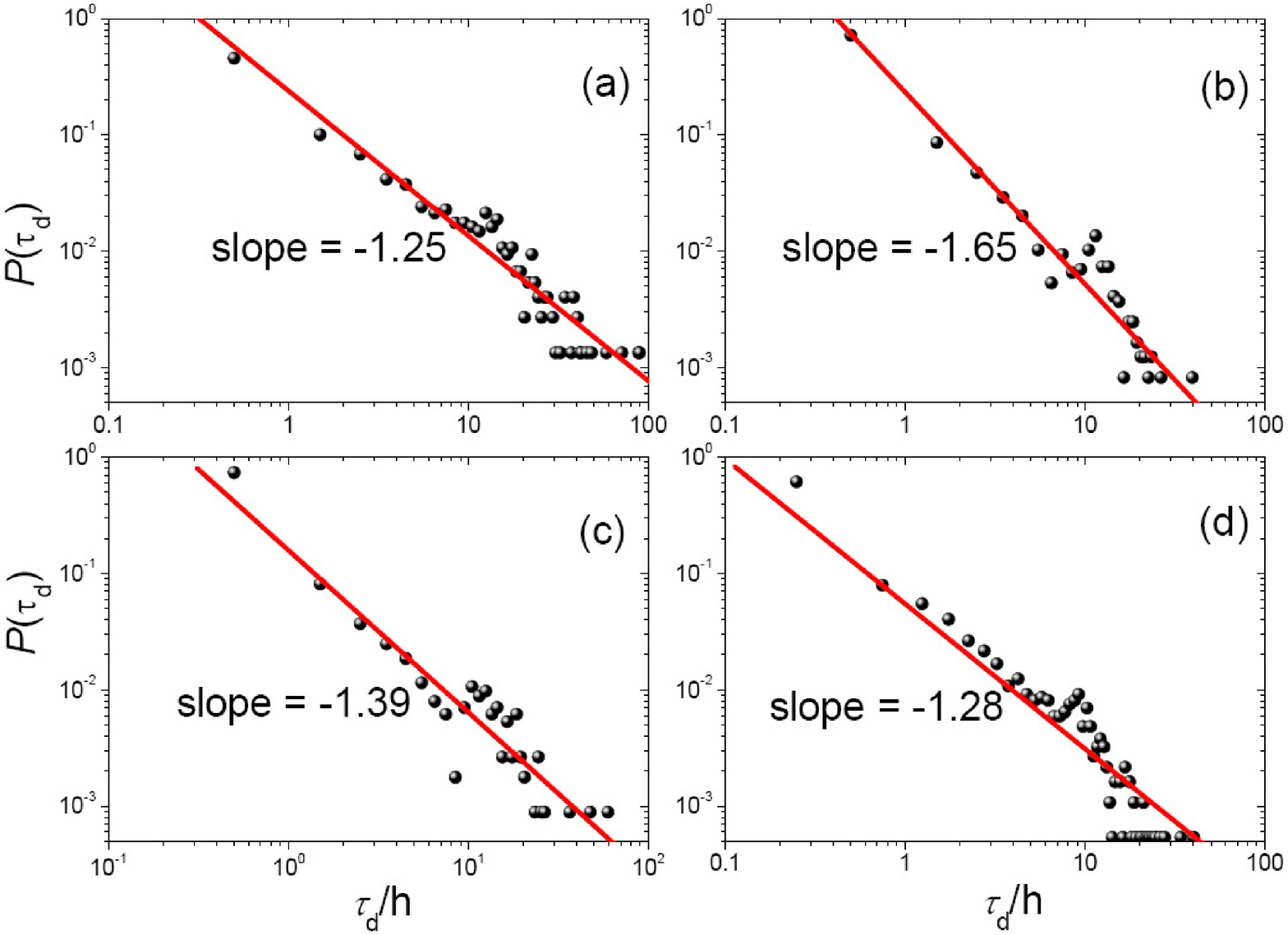}\\
  \caption{(color online)
The interevent time distributions of conversations in log-log plots.
The black dots represent empirical data, and the red lines are the
linear fittings. The panels (a), (b), (c) and (d) correspond to the
records of A, C, E and F, respectively.}
\end{figure}

\begin{figure}
  \includegraphics[width=8.5cm]{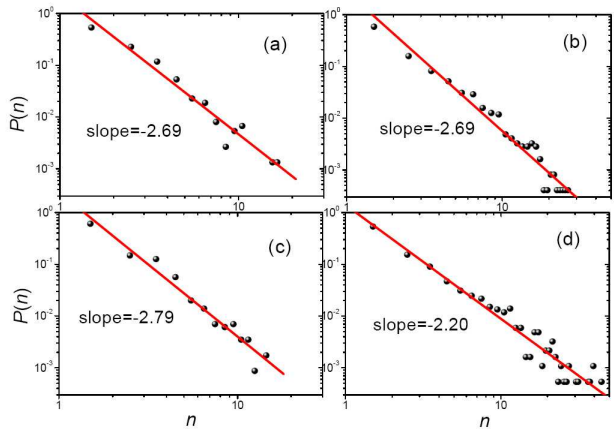}\\
  \caption{(color online)
The log-log plots of length distributions of conversations. The
black dots represent empirical data, and the red lines are the
linear fittings. The panels (a), (b), (c) and (d) correspond to the
records of A, C, E and F, respectively.}
\end{figure}

\section{CONCLUSION AND DISCUSSION}
In this Brief Report, we have investigated some statistical
properties of SM communication in the individual level. The
empirical evidence indicate that the SM communication pattern is
governed by a non-Poisson statistics. For commercial reason and the
right of personal privacy, we could not freely and automatically
download the SM data without permission. Therefore, those results
are limited by the lack of data. In despite of the small number of
samples, all the records display very similar statistics, thus we
believe those findings shown in this Brief Report are common for the
most of SM users.

The temporal statistics in SM communication are similar to those
observed in e-mail \cite{r18} and surface mail \cite{r19}
communications. Besides their similarity, the SM communication has
some specific features different from e-mail and surface mail.
Firstly, it is directly perceived through the senses to treat
surface mails as some tasks waiting for reply. We may soon reply the
urgent and important letters, and the ones not important or
difficult to reply may wait for a long time before being replied. In
contrast, we usually reply a short message immediately. It is seldom
seen that a short message is very hard to reply, thus we have to
spend a long time in preparing the response. Therefore, instead of
the HPF protocol \cite{r18}, the heavy-tailed distribution may root
in some other mechanisms, such as the competition with other tasks
\cite{r25}, the personal interest \cite{r26}, the social
interactions among many users \cite{net}, and so on. Secondly, a
single short message is usually a tiny part of a complete
conversation, thus to study the statistics in the resolution of
conversation may be more proper to reflect the function of SM
communication.

Based on the analytical solution \cite{Vazquez2005} of the
Barab\'asi model \cite{r18}, V\'azquez \emph{et al.} \cite{r25}
claimed the existence of two discrete universality classes of human
dynamics, whose characteristic power-law exponents are 1 and 1.5,
respectively. The e-mail communication, web browsing and library
loans belong to the former, while the surface mail communication
belongs to the latter. However, thus far, there are increasing
empirical evidence against the hypothesis of universality classes
for human dynamics \cite{Zhou2008}. As shown in Fig. 1, different
individual has different power-law exponents which are, typically,
larger than 1.5. Furthermore, Fig. 2 show a positive correlation
between activity and power-law exponent, which is against the
discrete universality classes. However, we also note that the
power-law exponents in the interevent distributions of conversations
are closer to 1.5. A clearer picture asks for abundant data in the
future.

\begin{acknowledgements}
The authors acknowledge Shuang-Xing Dai and Guan-Xiong Chen for
valuable discussion, as well as Ling-Zhi Hu and Kai Pan have for the
assistance of manuscript preparation. This work is partially
supported by the National Natural Science Foundation of China (Grant
Nos. 10472116 and 10635040), as well as the 973 Program under Grant
No. 2006CB705500.
\end{acknowledgements}


\begin{thebibliography}{ref1}
\bibitem{Watts2007} D.J. Watts, Nature {\bf 445}, 489 (2007).
\bibitem{r1} J.-P. Eckmann, E. Moses, and D. Sergi, Proc. Natl. Acad. Sci. U.S.A. {\bf 101}, 14333 (2004).
\bibitem{r2} W. Aiello, F. Chung, and L. Lu, Proceedings of the 32nd ACM Symposium on the Theory of Computing (pp. 171, ACM, New York, 2000).
\bibitem{r18} A.-L. Barabasi, Nature {\bf 435}, 207 (2005).
\bibitem{r19} J. G. Oliveira, and A.-L. Barab\'{a}si, Nature {\bf 437}, 1251 (2005).
\bibitem{Zhou2008} T. Zhou, X.-P. Han, and B.-H. Wang, arXiv: 0801.1389.
\bibitem{r20} V. Plerou, P. Gopikrishnan, L. A. N. Amaral, X. Gabaix, and H. E. Stanley, Phys. Rev. E {\bf 62}, 3023 (2000).
\bibitem{r21} J. Masoliver, and M. Montero, Phys. Rev. E {\bf 67}, 021112 (2003).
\bibitem{r22} T. Henderson, and S. Nhatti, Proc. 9th ACM Int. Conf. on Multimetia (pp. 212, ACM Press, 2001).
\bibitem{r23} T. Zhou, H. A.-T. Kiet, B. J. Kim, B.-H. Wang, and P. Holme, arXiv: 0711.4168.
\bibitem{r24} Z. Dezs\"{o}, E. Almaas1, A. Luk\'{a}cs, B. R\'{a}cz, I. Szakad\'{a}t, and A.-L. Barab\'{a}si, Phys. Rev. E {\bf 73}, 066132 (2006).
\bibitem{r25} A. V\'{a}zquez, J. G. Oliveira, Z. Dezs\"{o}, K.-I. Goh, I. Kondor, and A.-L. Barab\'{a}si, Phys. Rev. E {\bf 73}, 036127 (2006).
\bibitem{r26} X.-P. Han, T. Zhou, and B.-H. Wang, arXiv: 0711.0741.
\bibitem{r27} A. V\'{a}zquez, B. R\'acz, A. Luk\'{a}cs, and A.-L. Barab\'{a}si, Phys. Rev. Lett. {\bf 98}, 158702 (2007).
\bibitem{net} J. G. Oliveira, and A. V\'{a}zquez, arXiv: 0710.4916.
\bibitem{Vazquez2005} A. V\'{a}zquez, Phys. Rev. Lett. {\bf 95},
248701 (2005).

\end{thebibliography}
\end{document}